# Robust Adaptive Control of STATCOMs to Mitigate Inverter-Based-Resource (IBR)-Induced Oscillations

Hui Yuan, Linbin Huang, Huisheng Gao, Jikui Xing, Di Zheng, Ruisheng Diao

*Abstract*—**The interaction among inverter-based resources (IBRs) and power networks may cause small-signal stability issues, especially in low short-circuit-level grids. Besides, integrating static synchronous compensators (STATCOMs) in a multi-IBR system for voltage support can deteriorate small-signal stability. However, it is still challenging to fully understand the impact mechanism of STATCOMs on IBR-induced oscillation issues and to design optimal STATCOMs' control for dampening these oscillation issues in a multi-IBR system due to complex system dynamics and varying operating conditions. To tackle these challenges, this paper proposes a novel method to reveal how STATCOMs influence IBR-induced oscillation issues in a multi-IBR system from the viewpoint of grid strength, which can consider varying operating conditions. Based on this proposed method, *critical* operating conditions are identified, wherein the system tends to be most unstable. Moreover, we demonstrate that robust small-signal stability issues of the multi-IBR system with STATCOMs can be simplified as that of multiple subsystems under *critical* operating conditions, which avoids exhaustive studies on many operating conditions with detailed system models. On this basis, an adaptive control-parameter design method is proposed for STATCOMs to ensure system robust stability under varying operating conditions. The proposed methods are validated on a modified IEEE39-node test system.**

## I. INTRODUCTION

The increasing penetration of renewable resources, commonly interfacing with AC grids through power inverters, is changing modern power system dynamics and challenging secure grid operations[1],[2]. Particularly, inverter-based resources (IBRs) with widely-used grid-following controls tracking grid frequency through phase-lock loops (PLLs) may cause sub/sup-synchronous oscillation issues due to strong interaction between fast dynamics of IBRs and grid network, especially in low short-circuit-level grids[3]-[10]. Besides, in wind farms and photovoltaic plants, static synchronous compensator (STATCOM) is commonly used for providing reactive power compensation. However, STATCOMs can interact with IBRs and even deteriorate IBR-induced oscillation issues[11]-[12]. For instance, in 2015, a sub-synchronous resonance event was recorded in wind farms with STATCOMs in Hami, China[11].

STATCOMs have complex interactions with IBRs through power networks, especially considering multiple STATCOMs and IBRs, and varying operating conditions. This makes it hard to understand how STATCOMs influence IBR-induced oscillation issues. To understand STATCOMs' impact mechanism, Ref. [11] established the sequence impedance of a wind farm with a STATCOM and evaluated the damping provided by the STATCOM. The impedance-based analysis method commonly uses Nyquist stability criterion for small-signal stability analysis, which is based on reduced single-input single-output (SISO) transfer function of the system with the assumption that it has no right-half-plane pole. However, this assumption is not always satisfied[13]. To avoid this issue, Ref. [12] proposed a generalized short-circuit ratio (gSCR)-based method from the viewpoint of grid strength to reveal how the interaction among STATCOMs and IBRs impacts small-signal stability in a multi-machine system under rated operating conditions. It is noteworthy that varying operating conditions change the system's equilibrium and thus influence small-signal stability. However, it is still unknown how varying operating conditions influence the interaction among STATCOMs and IBRs, and the system's small-signal stability.

To suppress IBR-induced oscillation issues, many previous efforts improved control strategies, which can be divided into two categories: control strategy design of IBRs[14] and control strategy design of additional devices (e.g., STATCOMs[11],[15]). For instance, Ref. [14] proposed a PLL-reshaping method to improve the small-signal stability of a single-IBR infinite-bus system. However, IBRs are commonly packaged, and it is hard to modify IBR's control in practical operations. In comparison, STATCOMs are "white-boxed" models, and it is more convenient to modify STATCOMs' control. For instance, Ref. [11] proposed an intelligent parameter design method for STATCOMs to mitigate resonance in wind farms, which is based on the gain margin and phase margin of the system's SISO transfer function. However, this method may be ineffective if the obtained SISO transfer function has right-half-plane poles. Ref. [15] proposed an enhancing-grid stiffness control strategy of STATCOMs, which shapes STATCOMs' impedance as inductances and improves the stability by increasing grid strength or short-circuit ratio (SCR). However, these previous works mainly focused on one particular operating condition, which cannot ensure robust small-signal stability of the system under varying operating conditions.

To robustly mitigate IBR-induced oscillations under varying operating conditions, this paper proposes an adaptive control parameter design method for STATCOMs. We first propose a

H. Yuan, H. Gao, and J. Xing are with college of electrical engineering, Zhejiang university, Hangzhou 310027, China (Email: Yuan_Hui@zju.edu.cn; gaohuisheng@zju.edu.cn; 22210065@zju.edu.cn);

L. Huang is with department of information technology and electrical engineering, ETH Zürich, 8092 Zürich, Switzerland (e-mail: linhuang@ethz.ch);

D. Zheng is with College of Mechanical and Electrical Engineering, China Jiliang University, China 310018, China (e-mail: di.zh@cjlu.edu.cn).

R. Diao is with the ZJU-UIUC Institute, Zhejiang University, Haining, China 314400 (*Corresponding author*, email: ruishengdiao@intl.zju.edu.cn)

grid-strength-based method to analyze the small-signal stability of the multi-IBR system with STATCOMs under varying operating conditions, which was the extension of our previous work for the multi-IBR system with STATCOMs under rated operating conditions [12]. Based on the proposed method, the impact mechanism of STATCOMs on IBR-induced oscillation issues is revealed from the viewpoint of grid strength. Moreover, based on the proposed method, we find *critical* operating conditions wherein the system tends to be most unstable; and we demonstrate that the small-signal stability of the original system is bounded by multiple subsystems. On these bases, we simplify robust small-signal stability issues of the multi-IBR system with STATCOMs under varying operating conditions as that of multiple subsystems under *critical* operating conditions, which avoids exhaustive studies on many operating conditions with detailed system models. Moreover, an adaptive control parameter design method for STATCOMs is proposed to ensure robust small-signal stability of established multiple subsystems under *critical* operating conditions, which is also robustly effective for the original system under varying operating conditions. Besides, the proposed control method can consider the scenario that IBRs are "black-boxed" models. The main contributions of this paper can be summarized as:

1) A grid-strength-based method is proposed to evaluate the small-signal stability of the multi-IBR system with STATCOMs under varying operating conditions. On this basis, we reveal the impact mechanism of STATCOMs on IBR-induced oscillation issues from the viewpoint of grid strength.

2) Based on the proposed grid-strength-based method, the robust small-signal stability issues of a multi-IBR system with STATCOMs under varying operating conditions are simplified as multiple subsystems under *critical* operating conditions. This can avoid exhaustive simulation studies on many operating conditions using detailed system-wide models.

3) Based on the $H_\infty$ control theory[16], an adaptive control parameter design method is proposed for STATCOMs to ensure robust small-signal stability of established multiple subsystems under *critical* operating conditions. It is verified that the proposed control method can also ensure robust small-signal stability of the original system under varying operating conditions, even though the IBRs are "black-box" models.

The rest of this paper is organized as follows. Section II introduces the dynamic model of the multi-IBR system with STATCOMs under varying operating conditions and focused issues. In Section III, a grid-strength-based method is proposed to reveal the impact of STATCOMs on IBR-induced oscillation issues considering varying operating conditions. Section IV simplifies the robust small-signal stability issues of the multi-IBR system with STATCOMs under varying operating conditions. Section V proposes an adaptive control parameter design method for STATCOMs. In Section VI, the effectiveness of the proposed methods is demonstrated by eigenvalue analysis and electromagnetic transient simulations. Section VII draws the conclusions.

## II. System Modeling and Problem Description

Let us consider a multi-IBR system with $k$ STATCOMs as shown in Fig. 1, where nodes 1~$n$ are connected with IBRs, node $n+m+1$ is connected to external grids (simplified as an ideal voltage source); the remaining nodes are passive nodes. Since actual STATCOMs are not directly connected to IBRs' terminal nodes, we assume that STATCOM1~STATCOM$k$ are connected to passive nodes $n+1$~$n+k$. To simplify the analysis, we assume that: STATCOMs are all applied with ac-voltage control (AVC), and control parameters of STATCOMs are identical; AVC of STATCOMs is not saturated, i.e., reactive current reference $I_{qsref}$ is in the range of [-1,1]; IBRs' terminal voltage remains nearly at 1 p.u. due to STATCOMs' voltage support; we mainly consider the changes of IBRs' active power outputs and STATCOMs' reactive current outputs under varying operating conditions, but network parameters and control parameters of IBRs and STATCOMs are fixed.

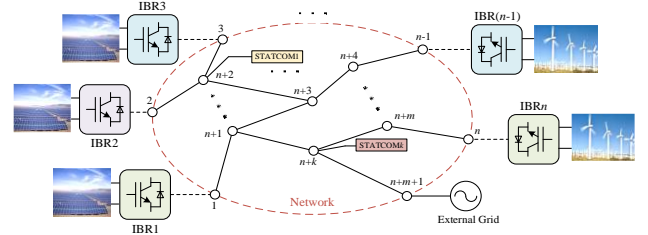

Fig. 1 One-line diagram of a typical multi-IBR system with STATCOMs.

When operating under non-rated conditions, the closed-loop characteristic equation of the multi-IBR system with STATCOMs can be given as Eq. (1) for small-signal stability analysis:

$$\det\left(\mathbf{Y}_{Gm}(s) + \mathbf{Y}_{Nm}(s)\right) = 0 \tag{1}$$

$$\mathbf{Y}_{Gm}(s) = \begin{bmatrix} \operatorname{diag}\left(\mathbf{Y}'_{IBRi}(s)\right) & \\ & \operatorname{diag}\left(\mathbf{Y}'_{STAj}(s)\right) \end{bmatrix} \tag{2}$$

$$\begin{cases} \mathbf{Y}'_{IBRi}(s) = P_{ei}\mathbf{Y}_{IBRi}(s), & i = 1,...,n \\ \mathbf{Y}'_{STAj}(s) = S_{Bsj}\mathbf{Y}_{STAj}(s), & j = 1,...,k \end{cases} \tag{3}$$

$$\mathbf{Y}_{STAj}(s) = \begin{bmatrix} Y_{11}(s) & I_{qsj}Y_{12}(s) \\ Y_{21}(s) & Y_{22}(s) \end{bmatrix} \tag{4}$$

$$\mathbf{Y}_{Nm}(s) = \mathbf{B}_{red} \otimes \boldsymbol{\gamma}(s) = \begin{bmatrix} \mathbf{B}_{11} & \mathbf{B}_{12} \\ \mathbf{B}_{21} & \mathbf{B}_{22} \end{bmatrix} \otimes \boldsymbol{\gamma}(s) \tag{5}$$

$$\boldsymbol{\gamma}(s) = \frac{1}{(s+\tau)^2/\omega_0 + \omega_0} \begin{bmatrix} (s+\tau) & \omega_0 \\ -\omega_0 & (s+\tau) \end{bmatrix} \tag{6}$$

where $\mathbf{Y}_{Gm}(s)$ and $\mathbf{Y}_{Nm}(s)$ denote admittance matrices of devices and network, wherein the similar derivation refers to Ref. [12]; diag() denotes a diagonal block matrix; $\mathbf{Y}'_{IBRi}(s)$ and $\mathbf{Y}'_{STAj}(s)$ are admittance matrices of IBR$i$ and STATCOM$j$, of which the detailed derivation is given in Ref. [17] and [18]; $P_{ei}$ and $S_{Bsj}$ are active power output of IBR$i$ and rated capacity of STATCOM$j$; $I_{qsj}$ is the reactive current output of STATCOM$j$; $Y_{11}(s)$, $Y_{12}(s)$, $Y_{21}(s)$, $Y_{22}(s)$ are elements of $\mathbf{Y}_{STAj}(s)$, which has no relation with $I_{qsj}$; $\mathbf{Y}_{IBRi}(s)$ and $\mathbf{Y}_{STAj}(s)$ are admittance matrices of IBR$i$ and STATCOM$j$ normalized at their rated capacities; $\mathbf{B}_{red} \in \mathbb{R}^{(n+k)\times(n+k)}$ is network node-reduced susceptance matrix containing IBR nodes and STATCOM nodes; $\mathbf{B}_{11} \in \mathbb{R}^{n\times n}$, $\mathbf{B}_{12} \in \mathbb{R}^{n\times k}$, $\mathbf{B}_{21} \in \mathbb{R}^{k\times n}$, $\mathbf{B}_{22} \in \mathbb{R}^{k\times k}$ are submatrices of $\mathbf{B}_{red}$; $\tau = R/L$ denotes line ratio of resistor ($R$) to inductance ($L$). We assume that all network lines have the same ratio $\tau$; $\omega_0$ is rated synchronous frequency.

By substituting Eqs. (2) and (5) into Eq. (1), the closed-loop characteristic equation Eq. (1) can be written as:

$$\det\left(\begin{bmatrix} \mathrm{diag}\left(P_{ei}\mathbf{Y}_{\mathrm{IBR}i}(s)\right) & \\ & \mathrm{diag}\left(S_{Bsj}\mathbf{Y}_{\mathrm{STA}j}(s)\right) \end{bmatrix} + \mathbf{B}_{red} \otimes \boldsymbol{\gamma}(s)\right) = 0 \quad (7)$$

It can be seen from Eq. (7) that IBRs interact with STATCOMs through the network, which may cause small-signal stability issues, especially in weak grids. To this end, this paper mainly focuses on addressing two main issues:

1) How to reveal the impact mechanism of STATCOMs on IBR-induced oscillation issues under varying operating conditions?

2) How to design effective control of STATCOMs to ensure the system's robust stability under varying operating conditions?

For *issue 1*, from Eq. (7), it is challenging to directly solve the characteristic equation for small-signal stability analysis because of complex interactions among IBRs and STATCOMs through the network. To deal with this issue, a grid-strength-based method is proposed in Section III. For *issue 2*, an adaptive parameter design method for STATCOMs is proposed in Section IV.

## III. Impact Mechanism of STATCOMs on IBR-Induced Oscillation Issues Under Varying Operating Conditions

### A. Proposed Grid-Strength-Based Method

In our previous work[12], a generalized short circuit ratio (gSCR)-based method is proposed to evaluate the small-signal stability of a multi-IBR system with STATCOMs under rated operating conditions. Specifically, the multi-IBR system with STATCOMs is converted into an equivalent homogeneous system for small-signal stability analysis. The equivalent homogeneous system can be further decoupled into $n$ independent subsystems for small-signal stability analysis, wherein these subsystems have the same equivalent device but different SCRs. Due to this, the small-signal stability of the equivalent homogeneous system (or the original system) can be represented by the *critical* subsystem with the smallest SCR (gSCR). The expression of gSCR is given as:

$$gSCR = \lambda_{\min}\left\{\mathbf{S}_B^{-1}\mathbf{B}_{redn}\right\}, \mathbf{S}_B = \mathrm{diag}\left\{S_{Bi}\right\}, \mathbf{B}_{redn} = \mathbf{B}_{11} - \mathbf{B}_{12}\mathbf{B}_{22}^{-1}\mathbf{B}_{21} \quad (8)$$

where $\lambda_{\min}\{\}$ denotes the smallest eigenvalue of a matrix; $\mathbf{S}_B$ is a diagonal matrix, wherein diagonal element $S_{Bi}$ is the rated capacity of IBR$i$; $\mathbf{B}_{11}$, $\mathbf{B}_{12}$, $\mathbf{B}_{21}$ and $\mathbf{B}_{22}$ refer to Eq. (5).

When control parameters of IBRs and STATCOMs are given, gSCR can evaluate the small-signal stability margin of the multi-IBR system with STATCOMs under rated operating conditions. The characteristic equation of the multi-IBR system with STATCOMs under rated operating conditions is given as:

$$\det\left(\begin{bmatrix} \mathrm{diag}\left(S_{Bi}\mathbf{Y}_{\mathrm{IBR}i}(s)\right) & \\ & \mathrm{diag}\left(S_{Bsj}\mathbf{Y}_{\mathrm{STA}j}(s)\right) \end{bmatrix} + \mathbf{B}_{red} \otimes \boldsymbol{\gamma}(s)\right) = 0 \quad (9)$$

It can be seen from Eqs. (7) and (9) that the multi-IBR system with STATCOMs under non-rated operating conditions has a similar characteristic equation as the system under rated operating conditions. Therefore, referring to [12], the multi-IBR system with STATCOMs under non-rated operating conditions can also be represented by an equivalent homogeneous system and its decoupled *critical* subsystem for small-signal stability analysis, as shown in Fig. 2(a). According to Ref. [12], the characteristic equation of the *critical* subsystem is given as:

$$\det\left(\mathbf{C}_1(s)\right) = \det\left(\overline{\mathbf{Y}}_S(s)\boldsymbol{\gamma}^{-1}(s) + \lambda_1\mathbf{I}_1\right) = 0 , \ \lambda_1 = \lambda_{\min}\left\{\mathbf{P}_e^{-1}\mathbf{B}_{redn}\right\} \quad (10)$$

$$\overline{\mathbf{Y}}_S(s) = \overline{\mathbf{Y}}_{\mathrm{IBR}}(s) + \overline{\mathbf{Y}}_{\mathrm{STA}}(s) \quad (11)$$

$$\overline{\mathbf{Y}}_{\mathrm{IBR}}(s) = \sum_{i=1}^{n} p_{1i}\mathbf{Y}_{\mathrm{IBR}i}(s), \overline{\mathbf{Y}}_{\mathrm{STA}}(s) = \sum_{j=1}^{k} p_{2j}\mathbf{Y}_{\mathrm{STA}j}(s) \quad (12)$$

$$p_{1i} = v_{1i}u_{1i} , \quad p_{2j} = S_{Bsj}\overline{\boldsymbol{u}}_1^T\begin{bmatrix} \mathbf{P}_e^{-1} & \\ & \mathbf{I}_k \end{bmatrix}\mathbf{E}_{sj}\overline{\boldsymbol{v}}_1 \quad (13)$$

$$\overline{\boldsymbol{u}}_1^T = \begin{bmatrix} \boldsymbol{u}_1^T & -\boldsymbol{u}_1^T\mathbf{P}_e^{-1}\mathbf{B}_{12}\mathbf{B}_{22}^{-1} \end{bmatrix}, \ \overline{\boldsymbol{v}}_1 = \begin{bmatrix} \boldsymbol{v}_1 \\ -\mathbf{B}_{22}^{-1}\mathbf{B}_{21}\boldsymbol{v}_1 \end{bmatrix} \quad (14)$$

where $\mathbf{C}_1(s)$ is the closed-loop transfer function matrix; $\overline{\mathbf{Y}}_S(s)$ is device's dynamics, which is a weighted sum of all IBRs and STATCOMs; $\mathbf{P}_e$=diag($P_{ei}$) is a diagonal matrix, wherein the diagonal element $P_{ei}$ is the active power output of IBR$i$; $p_{1i}$ ($i$=1,…,$n$) and $p_{2j}$($j$=1,…,$k$) are participation factors; $u_{1i}$ and $v_{1i}$ are $i$th elements of $\boldsymbol{u}_1^T$ and $\boldsymbol{v}_1$; $\boldsymbol{u}_1^T$ and $\boldsymbol{v}_1$ are normalized left and right eigenvectors of the smallest eigenvalue $\lambda_1$ for $\mathbf{P}_e^{-1}\mathbf{B}_{redn}$; $\mathbf{E}_{sj}$ is a square matrix representing the location of STATCOM$j$, wherein only ($n$+$j$)th diagonal element is one, and the other elements are zero. $\overline{\mathbf{Y}}_{\mathrm{STA}}(s)$ is rewritten as Eq. (15) under the assumption that STATCOMs have same control parameters:

$$\overline{\mathbf{Y}}_{\mathrm{STA}}(s) = p_{\Sigma}\overline{\mathbf{Y}}_{\mathrm{STA1}}(s), \overline{\mathbf{Y}}_{\mathrm{STA1}}(s) = \begin{bmatrix} Y_{11}(s) & I_{q\Sigma}\tilde{Y}_{12}(s) \\ Y_{21}(s) & Y_{22}(s) \end{bmatrix} \quad (15)$$

$$p_{\Sigma} = \sum_{j=1}^{k} p_{2j}, I_{q\Sigma} = \sum_{j=1}^{k} p_{2j}I_{qsj}\Bigg/ p_{\Sigma} \quad (16)$$

From Eqs. (15) and (4), $\overline{\mathbf{Y}}_{\mathrm{STA}}(s)$ can be considered as the dynamic of a STATCOM, where $p_{\Sigma}$ and $I_{q\Sigma}$ represent capacity and reactive current output. Moreover, since $p_{2j}{\geq}0$ and $I_{qsj}$ is in the range of [-1,1], $I_{q\Sigma}$ is in the range of [-1,1]. Thus, the dynamics of $\overline{\mathbf{Y}}_S(s)$ in Eq. (11) can be considered as the weighted sum of IBRs and a STATCOM, as shown in Fig. 2(a).

From Eqs. (8) and (10), it can be seen that the gSCR and $\lambda_1$ have similar forms. In other words, the gSCR can be extended to evaluate the small-signal stability of the multi-IBR system with STATCOMs under non-rated operating conditions when control parameters of IBRs and STATCOMs are given. The difference is that the gSCR is minimal eigenvalue of $\mathbf{P}_e^{-1}\mathbf{B}_{redn}$ for varying operating conditions, i.e.,

$$gSCR = \lambda_{\min}\left\{\mathbf{P}_e^{-1}\mathbf{B}_{redn}\right\} \quad (17)$$

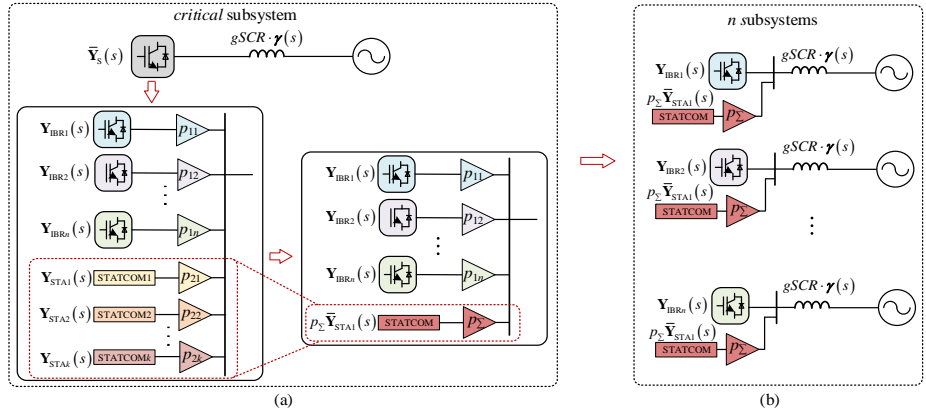

Fig. 2 (a) *critical* subsystem representing small-signal stability of the multi-IBR system with STATCOMs under varying operating conditions; (b) *n* subsystems that encircle the small-signal stability of the *critical* subsystem as introduced in Section IV.B.

Eq. (17) can also consider rated operating conditions. In this case, IBR*i*'s active power output is equal to the rated capacity(i.e., $S_{Bi}$), and thus Eq. (17) is equivalent to Eq. (8).

The *critical* gSCR (i.e., CgSCR) is given as, which represents that the system is *critically* stable:

$$CgSCR = \underset{gSCR}{\arg}\left\{\det\left(\tilde{\mathbf{Y}}_S\left(s_c\right)\boldsymbol{\gamma}^{-1}\left(s_c\right)+gSCR\cdot\mathbf{I}_1\right)\right\}=0 \qquad (18)$$

where arg{ } denotes the eigenvalue calculation of CgSCR; $s_c=j\omega_c$ is dominant eigenvalues of the *critical* subsystem in Eq. (10); $\omega_c$ is oscillation frequency.

Based on the gSCR and CgSCR, we can evaluate small-signal stability of the multi-IBR system with STATCOMs under varying operating conditions from the viewpoint of grid strength. That is, if gSCR>CgSCR, the system is stable; the larger the (gSCR-CgSCR) is, the more stable the system is; otherwise, if gSCR<CgSCR, the system is unstable.

### B. Analysis of STATCOMs' Impact on Small-Signal Stability

As discussed in the above subsection, the multi-IBR system with STATCOMs under varying operating conditions can be represented by the *critical* subsystem Eq. (10) for small-signal stability analysis, which considers the change of IBRs' active power outputs $P_{ei}$, ($i=1,...,n$) and STATCOMs' reactive current outputs $I_{qsj}$, ($j=1,...,k$). Moreover, when control parameters of IBRs and STATCOMs are given, the grid-strength indicator (i.e., gSCR) can evaluate the small-signal stability margin of the *critical* subsystem (or the original system) under varying operating conditions. Based on the *critical* subsystem and grid-strength indicator, the impact mechanism of STATCOMs on the system stability under varying operating conditions can be described as follows:

1) The connection of STATCOMs has no impact on gSCR, but has impacts on CgSCR. This is because gSCR in *critical* subsystem is that of the multi-IBR system without STATCOMs as shown in Eq. (17), but the calculation of CgSCR is related to STATCOMs' dynamics as shown in Eq. (18).

2) STATCOMs can increase or decrease CgSCR, which is mainly related with the dynamics of STATCOMs referring to Eqs. (12) and (18), including control parameters and reactive current outputs $I_{qsj}$. The detailed discussion refers to Section VI. Besides, when STATCOMs decrease CgSCR with fixed gSCR, the system becomes more stable; otherwise, if STATCOMs increase CgSCR with fixed gSCR, the system tends to be unstable. This is because the (gSCR-CgSCR) represents the system's stability margin, and the larger (gSCR-CgSCR) is, the system is more stable as discussed in the above subsection.

3) Participation factor $p_{2j}$ reflects the relative degree of the impact of STATCOM*j* on CgSCR or (system's stability). We can see from Eqs. (11), (12) and (18) that the larger $p_{2j}$ is, the larger the impact of STATCOM*j* on CgSCR is.

4) The decrease of IBRs' active power outputs $P_{ei}$, ($i=1,...,n$) increases participation factor $p_\Sigma$ in Eq. (16) and thus increases the relative degree of STATCOMs' impact on CgSCR. To be specific, since STATCOMs are commonly connected near IBRs, it is reasonable to assume that STATCOMs are connected to IBR nodes. Due to this, $p_\Sigma$ in Eq. (16) is written as

$$p_\Sigma = \sum_{i=1}^{n}\frac{S_{Bsi}}{P_{ei}}\,p_{1i} \qquad (19)$$

where $S_{Bsi}$ is capacity of STATCOM*i*, and satisfies $S_{Bsi}\geq0$, wherein $S_{Bsi}=0$ means there is no STATCOMs near IBR*i*;

Since participation factor $p_{1i}$ satisfies $0<p_{1i}<1$ and $\sum_{i=1}^{n}p_{1i}=1$, we can conclude from Eq. (19) that the decrease of IBRs' active power outputs $P_{ei}$, ($i=1,...,n$) commonly increase $p_\Sigma$. We mention that the decrease of $P_{ei}$ not only influences CgSCR, but also increases gSCR. This is because the sensitivity of gSCR for IBRs' active power outputs $P_{ei}$, ($i=1,...,n$) is negative[19].

5) The decrease of STATCOMs' reactive current outputs $I_{qsj}$, ($j=1,...,k$) only influences CgSCR, but has no impact on gSCR. As shown in Eqs. (15) and (16), the change of $I_{qsj}$, ($j=1,...,k$) mainly influences STATCOM's dynamics $\tilde{\mathbf{Y}}_{STA1}(s)$ and thus influences CgSCR. Besides, referring to the above 1), we can conclude that $I_{qsj}$, ($j=1,...,k$) have no impact on gSCR. Moreover, it is worth noting that whether the increase of STATCOMs' reactive current outputs $I_{qsj}$ increases or decreases CgSCR is uncertain, which depends on STATCOMs' dynamics. This will be discussed in Section VI.

## IV. SIMPLIFYING SYSTEM ROBUST SMALL-SIGNAL STABILITY ISSUES UNDER VARYING OPERATING CONDITIONS

In this section, we will first discuss *critical* operating conditions in that the system tends to be most unstable. Then,

we will furtherly demonstrate that the small-signal stability of the *critical* subsystem (representing the stability of the original system as discussed in Section III.A) is bounded by that of $n$ subsystems. On these bases, the robust small-signal stability issue of the multi-IBR system with STATCOMs under varying operating conditions can be simplified as that of multiple subsystems under *critical* operating conditions, which will be used in Section V.

### A. Discussion of Critical Operating Conditions

When the operating condition changes, so do IBRs' active power outputs $P_{ei}$, $(i=1,...,n)$ and STATCOMs' reactive currents outputs $I_{qsj}$, $(j=1,...,k)$. According to 3)~5) in Section III.B, we can see that: the decrease of $P_{ei}$ increases $p_\Sigma$ and gSCR; $I_{qsj}$ only impacts CgSCR but has no impact on gSCR; besides, the impact of $I_{qsj}$ on CgSCR is uncertain, which depends on STATCOMs' dynamics. On these bases, we give a proposition about *critical* operating conditions.

**Proposition 1:** If STATCOMs decrease CgSCR, then the *critical* operating conditions are that all IBRs output rated active power (i.e., $S_{Bi}$, $i=1,...,n$) with STATCOMs' reactive current output $I_{qsj}$ in the range of [-1, 1].

*Proof:* When STATCOMs decrease CgSCR, the decrease of IBRs' active power outputs causes the decrease of CgSCR due to that $p_\Sigma$ increases. Besides, since the decrease of IBRs' active power output increases gSCR, (gSCR-CgSCR) becomes larger (i.e., the system becomes more stable) with the decrease of IBRs' active power outputs. That is, if the connection of STATCOMs decreases CgSCR, the *critical* subsystem tends to be most unstable under *critical* operating conditions. The proof is concluded.

In other words, *critical* operating conditions can be considered as that $gSCR=\lambda_{\min}\left\{\mathbf{S}_B^{-1}\mathbf{B}_{rede}\right\}$ in Eq. (8) and $I_{q\Sigma}$ in the range of [-1,1] for the *critical* subsystem with the pre-condition that STATCOMs decrease CgSCR. Note that this pre-condition can be satisfied by the proposed control method in Section V.

### B. Multi-Subsystem Encircling Stability of Critical Subsystem

The small-signal stability of the multi-IBR system with STATCOMs under varying operating conditions is bounded by the dynamics of IBRs and STATCOMs. To analyze the small-signal stability of the multi-IBR system with STATCOMs, we define $n$ equivalent subsystems as shown in Fig. 2(b). These $n$ equivalent subsystems have different device dynamics but the same gSCR. In each subsystem, the devices include a IBR in the original system and a STATCOM $\overline{\mathbf{Y}}_{STA}(s)$ in Eq. (15). These two devices are parallel in each subsystem as shown in Fig. 2 (b). Besides, the small-signal stability margin of these $n$ subsystems is ranked in the order from smallest to largest, i.e., subsystem$_{1,1}$, subsystem$_{1,2}$, ..., and subsystem$_{1,n}$.

From Fig. 2 (a), the participation factor $p_{1i}$ $(i=1,...,n)$ determines the stability of the *critical* subsystem with given gSCR and dynamics of IBRs and STATCOMs. That is, if the participation factor $p_{1i}$ of the IBR's dynamic (same as that of the subsystem$_{1,1}$) is larger, the *critical* subsystem will tend to be unstable; if the participation factor $p_{1i}$ of the IBR's dynamic

(same as that of the subsystem$_{1,n}$) is larger, the critical subsystem will be more stable. Since the participation factor $p_{1i}$ $(i=1,...,n)$ satisfies $0<p_{1i}<1$ $(i=1,...,n)$, the extreme cases for the *critical* subsystem with given gSCR and dynamics of IBRs and STATCOMs are that: 1) when participation factor $p_{1i}$ of IBR's dynamic (same as that of subsystem$_{1,1}$) is one, and the other $p_{1i}$ are all zero (i.e., subsystem$_{1,1}$), the *critical* subsystem is most unstable; 2) when $p_{1i}$ of IBR's dynamic (same as that of subsystem$_{1,n}$) is one, and the other $p_{1i}$ are all zero (i.e., subsystem$_{1,n}$), the *critical* subsystem is most stable.

As a result, the small-signal stability of the *critical* subsystem (or the multi-IBR system with STATCOMs) is bounded by subsystem$_{1,1}$ and subsystem$_{1,n}$, which will be illustrated in Section VI. Besides, since the system tends to be most unstable under *critical* operating conditions, as discussed in the above subsection, we can simplify the robust small-signal stability issue of the multi-IBR system with STATCOMs under varying operating conditions as that of subsystem$_{1,1}$~ subsystem$_{1,n}$ under *critical* operating conditions.

## V. Adaptive Parameter Design of STATCOMs Under Varying Operating Conditions

As discussed in Section IV, the control design issue of the multi-IBR system with STATCOMs for robust small-signal stability under varying operating conditions can be simplified as that of subsystem$_{1,1}$~subsystem$_{1,n}$ under *critical* operating conditions. Besides, since IBRs are commonly "black-boxed" and thus it is hard to modify IBRs' control strategies in practical operations, we intend to modify STATCOMs' control strategies, which is more convenient. In this section, an adaptive control parameter design method for STATCOMs is proposed to ensure robust small-signal stability of subsystem$_{1,1}$~subsystem$_{1,n}$ under *critical* operating conditions.

### A. Adaptive Control Parameter Design Method for STATCOMs

$\mathcal{H}_\infty$-synthesis design is an efficient way to ensure robust small-signal stability, which will be used for adaptive control-parameter design of STATCOMs. For brevity, we omit the detailed introduction of $\mathcal{H}_\infty$-synthesis design, which can be found in Ref. [16]. Based on $\mathcal{H}_\infty$-synthesis method and the analysis in Section IV, the adaptive control parameter design problem of STATCOMs is described as simultaneous stabilization problems of subsystem$_{1,1}$~subsystem$_{1,n}$ under *critical* operating conditions or a min-max optimization problem:

$$\min_{\mathbf{K}}\ \max_{\substack{i=1,...,n,\ I_{q\Sigma}\in[-1,1]\\ gSCR=\lambda_{\min}\left(\mathbf{S}_B^{-1}\mathbf{B}_{rede}\right)}}\left\|\mathbf{C}_i\left(s\mathbf{I}-\left(\mathbf{A}_i+\mathbf{B}_i\mathbf{K}\mathbf{C}_i\right)\right)^{-1}\mathbf{B}_i\right\|_\infty \quad (20)$$

where $\|.\|_\infty$ is the infinite norm; $\mathbf{C}_i(s\mathbf{I}-(\mathbf{A}_i+\mathbf{B}_i\mathbf{K}\mathbf{C}_i))^{-1}\mathbf{B}_i$ is closed-loop transfer function matrix of subsystem$_{1,i}$ under *critical* operating conditions; $\mathbf{A}_i$, $\mathbf{B}_i$, $\mathbf{C}_i$ are parameter matrices of the open-loop state-space model of subsystem$_{1,i}$, described as:

$$\text{subsystem}_{1,i}:\ \begin{cases}\dot{\mathbf{x}}_i=\mathbf{A}_i\mathbf{x}_i+\mathbf{B}_i\mathbf{u}_i\\ \mathbf{y}_i=\mathbf{C}_i\mathbf{x}_i\end{cases} \quad (21)$$

where $\mathbf{x}_i$, $\mathbf{y}_i$, and $\mathbf{u}_i$ are vectors of state variables, algebraic variables, and input variables. Note that if IBRs are "black-boxed", parameter matrices $\mathbf{A}_i$, $\mathbf{B}_i$ and $\mathbf{C}_i$ can be obtained by identifying internal dynamics through rational approximation based on frequency scan[20]. $\mathbf{K}$ in Eq. (20) represents the $\mathcal{H}_\infty$-controller in STATCOMs, which can be a transfer function matrix related with "$s$", or a static gain matrix. To simplify the difficulty of the adaptive parameter control design of STATCOMs, we do not change the original control structure of STATCOMs, and choose proportional-integral (PI) parameters in AVC and PLL of STATCOMs as elements in $\mathbf{K}$. That is, matrix $\mathbf{K}$ can be expressed as:

$$\boldsymbol{u}_i = \mathbf{K}\boldsymbol{y}_i, \mathbf{K} = \begin{bmatrix} k_{acps} & k_{acis} & 0 & 0 \\ 0 & 0 & k_{pllps} & k_{pllis} \end{bmatrix} \quad (22)$$

where $\boldsymbol{y}_i=[U_{qsi}, x_{1si}, (U_{si}-1), x_{2si}]^{\mathrm{T}}$, wherein $U_{qsi}$ and $U_{si}$ are $q$-axis components and amplitude of STATCOM's terminal voltage in subsystem$_{1,i}$; $x_{1si}$ and $x_{2si}$ are the integrations of $U_{qsi}$ and $(U_{si}-1)$; $\boldsymbol{u}_i=[\omega_{pllsi}, I_{qrefsi}]^{\mathrm{T}}$, wherein $\omega_{pllsi}$ is PLL's frequency output, and $I_{qrefsi}$ is $q$-axis current reference; $k_{acps}$ and $k_{acis}$ are PI parameters in AVC; $k_{pllps}$ and $k_{pllis}$ are PI parameters in PLL.

The optimization problem in Eq. (20) is a pure-stabilization $\mathcal{H}_\infty$-synthesis problem, of which the necessary and sufficient conditions for the system's robust small-signal stability is that the objective function in Eq. (20) is finite[16]. Besides, if the objective function in Eq. (20) is smaller, the obtained control parameters of STATCOMs can improve the system's stability better. However, the range of STATCOM's reactive current output $I_{q\Sigma}$ (i.e., [-1, 1]) is large in *critical* operating conditions. Due to this, the optimization problem Eq. (20) may have no solutions, which will be discussed in Section VI. To deal with this issue, we divide *critical* operating conditions as $m$ sub-conditions. In these sub-conditions, $gSCR= \lambda_{\min}\{\mathbf{S}_B^{-1}\mathbf{B}_{redn}\}$ and $I_{q\Sigma}$ is in $m$ intervals, i.e., [-1, -1+2/m), [-1+2/m, -1+4/m),…,[-1+2(m-1)/m, 1]. For each sub-condition, we establish a corresponding optimization problem given in Eq. (20) and obtain the corresponding controller $\mathbf{K}_i$ in Eq. (22) to ensure robust small-signal stability of the $n$ subsystems.

In other words, to ensure robust small-signal stability of the multi-IBR system with STATCOMs under varying operating conditions, we firstly obtain a set of controllers $\{\mathbf{K}_1,…,\mathbf{K}_m\}$ for STATCOMs by off-line solving established $m$ optimization problems Eq. (20), and then on-line adaptively choose proper controller $\mathbf{K}_i$ according to real-time STATCOMs' reactive current outputs and IBRs' active power outputs. Note that by experiment, we find that the obtained optimal controller $\mathbf{K}_i$ can ensure to decrease CgSCR of the *critical* subsystem under *critical* operating conditions; besides, if $m$ is larger, $\mathbf{K}_i$ can cause a larger decrease of CgSCR (or improve the system's stability better); but if $m$ is too large, it will cause a high demand of real-time communication and computation for $I_{q\Sigma}$; if $m$ is too small or even equal to 1, the optimization problem in Eq. (20) may have no solutions, which will be illustrated in Section VI. Thus, $m$ should be properly chosen to balance control performance and economic cost.

### B. Implementation Procedure of Proposed Control Method

As discussed above, the proposed adaptive control parameter design method for STATCOMs in the multi-IBR system with STATCOMs under varying operating conditions mainly includes two parts: 1) Off-line calculate control-parameter set $\{\mathbf{K}_1,…,\mathbf{K}_m\}$ for STATCOMs by solving established $m$ optimization problems in Eq. (20), which is aimed to ensure robust small-signal stability of subsystem$_{1,1}$~subsystem$_{1,n}$ under $m$ sub-conditions; 2) On-line adjust STATCOMs' control parameters based on real-time STATCOMs' reactive current outputs, and IBRs' active power outputs, which assumes that real-time STATCOMs' reactive current outputs and IBRs' active power outputs can be obtained. The implementation procedure of this proposed adaptive control-parameter design method is shown in Fig. 3, with the main steps summarized as:

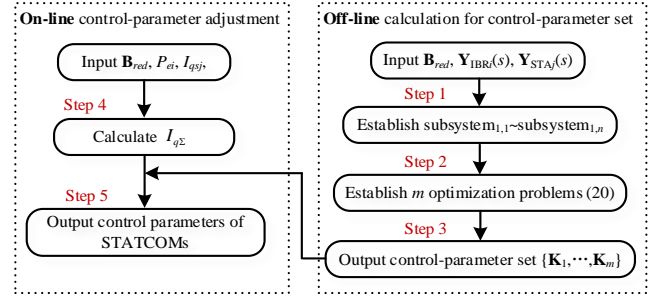

Fig. 3 Flowchart of the proposed adaptive control parameter design method for STATCOMs.

1) In the off-line process, establish subsystem$_{1,1}$ ~subsystem$_{1,n}$ for the multi-IBR system with STATCOMs under *critical* operating conditions. If the inner parameters of IBRs are unknown, their internal dynamics can be identified by a rational approximation based on frequency scan[20];

2) Establish $m$ optimization problems in Eq. (20) to ensure small-signal stability of subsystem$_{1,1}$~subsystem$_{1,n}$ under $m$ sub-conditions, wherein $m$ sub-conditions are obtained by dividing *critical* operating conditions. That is, $gSCR= \lambda_{\min}\{\mathbf{S}_B^{-1}\mathbf{B}_{redn}\}$ and $I_{q\Sigma}$ is in $m$ intervals, i.e., [-1, -1+2/m), [-1+2/m, -1+4/m),…,[-1+2(m-1)/m, 1];

3) Output control-parameter set $\{\mathbf{K}_1, …, \mathbf{K}_m\}$ obtained by solving $m$ optimization problems in Eq. (20) through Matlab solvers;

4) In the on-line process, calculate real-time reactive current output $I_{q\Sigma}$ in Eq. (16) based on real-time IBRs' active power outputs and STATCOMs' reactive power outputs;

5) On-line adjust control parameters of STATCOMs based on control-parameter set $\{\mathbf{K}_1, …, \mathbf{K}_m\}$ and real-time $I_{q\Sigma}$. To be specific, if $I_{q\Sigma}$ is in the interval [-1+2(i-1)/m, -1+2i/m), then control parameters of all STATCOMs are set as $\mathbf{K}_i$.

## VI. CASE STUDIES

In this section, the proposed grid-strength-based method and adaptive control method are validated by MATLAB/Simulink on a modified IEEE 39-node system as shown in Fig. 4, where nodes 1~9 are connected with IBRs through a set-up transformer, and node 39 is connected to external grids, simplified as an infinite bus. Each IBR represents a wind farm. IBRs' capacities refer to TABLE I. In practice, wind farms are commonly installed with a certain percentage of STATCOMs for voltage support. Due to this, we installed 30%-capacity STATCOM for each IBR located at the high-voltage side, as shown in Fig. 4. Network parameters refer to Ref. [21]. Control parameters of IBRs and STATCOMs refer to TABLEs. II and III.

### A. Verification of Theoretical Analysis in Sections III and IV

*1) Verification of proposed gSCR-based method.* As discussed in Section III, the theoretical analysis of STATCOMs' impact on the system's stability under varying operating conditions is based on the proposed gSCR-based method. Due to this, we first verify the validity of the proposed gSCR-based method by eigenvalue analysis in a modified IEEE39-node system, as shown in Fig. 4. In this system, all STATCOMs use the same AVC, but IBRs are heterogeneous: IBR1~IBR3 use constant active power control, and the other IBRs use dc voltage control, referring to TABLE II. Several cases are created by increasing active power outputs of all IBRs from 0.5 to 1 p.u at the same proportion, which are normalized at their rated capacities. The corresponding gSCR decreases from 3.36 to 1.68. Under these cases, we compare the resulting dominant eigenvalues of the modified 39-node system and its *critical* subsystem, as shown in Fig. 5 (a). The corresponding loci of the damping ratio of dominant eigenvalues for the modified 39-node system are given in Fig. 5 (b) represented by a red dotted curve.

larger (gSCR-CgSCR) is, the more stable the system is; otherwise, if gSCR<CgSCR, the system is unstable. The same conclusion can also be observed from the loci of the damping ratio of the system in Fig. 5 (b). Simulation results are consistent with the theoretical analysis in Section III.

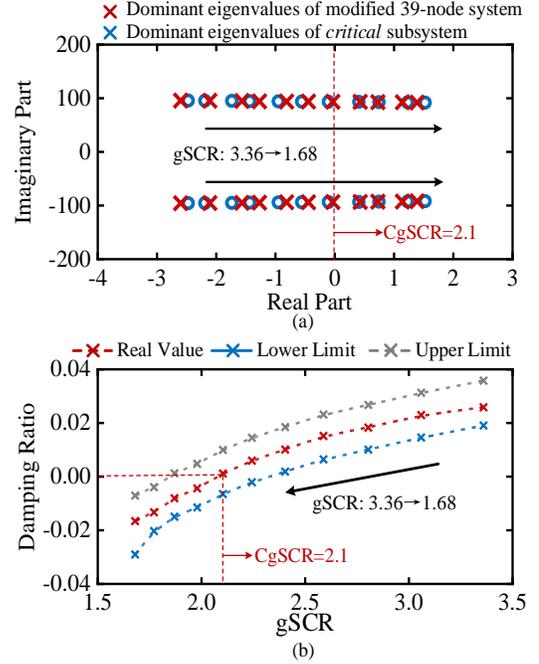

Fig. 5 when gSCR changes from 3.36 to 1.68: (a) Loci of dominant eigenvalues of the modified 39-node system and *critical* subsystem; (b) Loci of the damping ratio of dominant eigenvalues for modified 39-node system (red dotted curve), the weakest subsystem (blue dotted curve) and strongest subsystem ((grey dotted curve)) in subsystem$_{1,1}$ ~ subsystem$_{1,9}$.

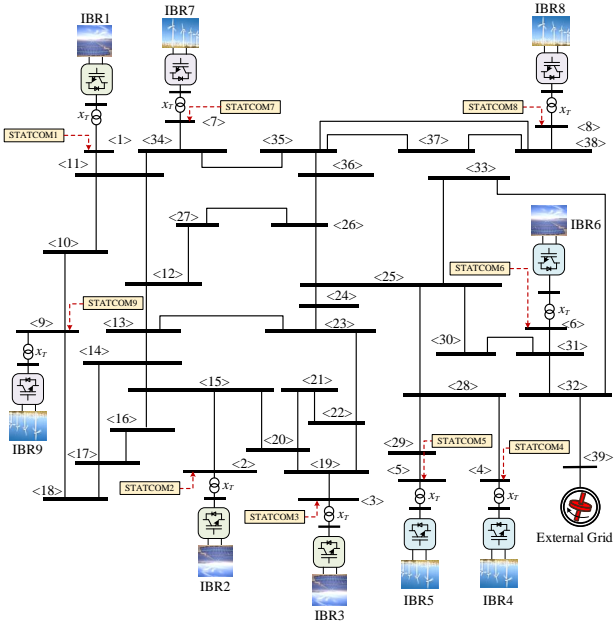

Fig. 4 A nine-IBR system with nine STATCOMs.

It can be seen from Fig. 5 (a) that the dominant eigenvalues of the *critical* subsystem match very well with the modified 39-node system when gSCR is changed from 3.36 to 1.68. This indicates that the small-signal stability of the multi-IBR system with STATCOMs can be represented by that of the *critical* subsystem under varying operating conditions.

Moreover, Fig. 5 (a) shows that gSCR can evaluate the small-signal stability of the modified 39-node system under varying operating conditions when control parameters of IBRs and STATCOMs are given. As shown in Fig. 5 (a), CgSCR is equal to 2.1. When gSCR=CgSCR, the system is critically stable; when gSCR>CgSCR, the system is stable; besides, the

#### TABLE I
#### RATED CAPACITIES OF EACH IBR (PER-UNIT)

| IBR1 | IBR2 | IBR3 | IBR4 | IBR5 |
|------|------|------|------|------|
| 1 | 2 | 1 | 1 | 2 |
| IBR6 | IBR7 | IBR8 | IBR9 | |
| 10 | 2 | 2 | 1 | |

#### TABLE II
#### CONTROL PARAMETERS OF IBRs (PER-UNIT)

| IBRs | The PLL $H_{pll}(s)$ | DC voltage loop $H_{dc}(s)$ | Constant active power control loop $H_{ff}(s)$ |
|------|----------------------|------------------------------|------------------------------------------------|
| 1~3 | 16+9500/s | / | 1+5/s |
| 4~6 | 13+9800/s | 0.5+5/s | / |
| 7~9 | 16+9500/s | 0.5+5/s | / |

Filter inductance $L_f$, filter capacitance $C_f$, dc capacitance $C_{dc}$: 0.05, 0.05, 0.038;
Transfer function of the current control loop $H_i(s)$: 1.5+10/s;
Transfer function of the voltage feedforward control loop $H_{ff}$: 1/(1+0.0001s);
$q$-axis current reference $I_{qref}$: 0.

#### TABLE III
#### CONTROL PARAMETERS OF STATCOMs (PER-UNIT)

Filter inductance $L_{fs}$, dc capacitance $C_{dcs}$: 0.1, 0.038;
Transfer function of current control loop $H_{is}(s)$: 1+10/s;
Transfer function of dc voltage control loop $H_{dcs}(s)$: 1+5/s;
Transfer function of the PLL $H_{plls}(s)$: 30+7000/s;
Transfer function of the AVC $H_{acs}(s)$: 1+5/s;

Furtherly, the electromagnetic time-domain simulation of the modified 39-node system is provided to verify the validity of the proposed gSCR-based method. When the active power outputs of all IBRs are set as 0.7 and 0.9 p.u., the corresponding gSCR are 2.404 and 1.87. Under these two cases, a disturbance is applied to the infinite bus at *t*=1*s* to cause 0.05 p.u. voltage drop and lasts 0.05s. Time-domain responses of active power outputs of all IBRs under these two cases are given in Fig. 6.

It can be seen from Fig. 6 that when gSCR=2.404>2.1(i.e., CgSCR in Fig. 5), the system is stable. However, when gSCR=1.87 <CgSCR, the system is unstable. Time-domain simulation results are consistent with eigenvalue results from Fig. 5, which verifies the effectiveness of the proposed gSCR-based method.

*2) Verification of STATCOMs' Impact Analysis.* As discussed in Section III.B, STATCOMs only influence CgSCR. Besides, STATCOMs' impact on CgSCR is uncertain, which depends on STATCOMs' dynamics. Here, we use the *critical* subsystem of the modified 39-node system (all IBRs output rated active power) to discuss how STATCOMs influence CgSCR. Five scenarios are considered:

1) Control parameters of STATCOM in the *critical* subsystem refer to TABLE III and $I_{q\Sigma}$=-0.5 p.u.;

2) Control parameters of STATCOM in the *critical* subsystem refer to TABLE III and $I_{q\Sigma}$=0.5 p.u.;

3) Control parameters of STATCOM in the *critical* subsystem refer to TABLE III, but PI parameters of AVC and PLL are set as "2.92, 5" and "10.3, 20000", and $I_{q\Sigma}$=0.5 p.u.;

4) Control parameters of STATCOM in the *critical* subsystem refer to TABLE III, $I_{q\Sigma}$=0.5 p.u., and $p_{\Sigma}$=0.4;

5) Disconnect STATCOM as a reference.

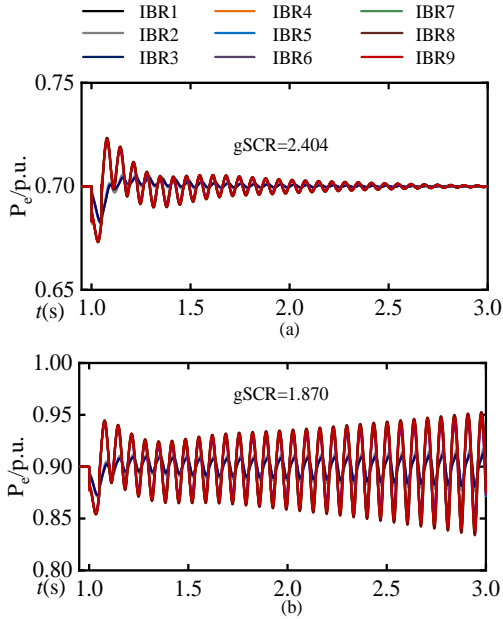

Fig. 6 Time-domain responses of active power outputs of IBR1~IBR9: (a) gSCR=2.404; (b) gSCR=1.87.

TABLE IV
CgSCR OF CRITICAL SUBSYSTEM UNDER SCENARIOS 1)~5)

|  | Scenario 1) | Scenario 2) | Scenario 3) | Scenario 4) | Scenario 5) |
|---|---|---|---|---|---|
| $p_{\Sigma}$ | 0.3 | 0.3 | 0.3 | 0.4 | 0 |
| CgSCR | 1.31 | 2.24 | 1.28 | 2.3 | 1.94 |

TABLE IV shows the CgSCR of the *critical* subsystem under these five scenarios. According to scenarios 1), 2), and 5), we can see that when $I_{q\Sigma}$ changes, STATCOM may increase (or decrease) CgSCR, i.e., deteriorate (or improve) the system's stability. We can see from scenarios 2), 3), and 5) that when STATCOM's control parameters change, STATCOM may also increase (or decrease) CgSCR. These demonstrate that STATCOM's impact on CgSCR is uncertain depending on STATCOM's dynamics, including control parameters and reactive current output. Besides, comparing scenarios 2), 4), and 5), we can see that the increase of $p_{\Sigma}$ enlarges CgSCR. This indicates that $p_{\Sigma}$ represents the relative degree of STATCOM's impact on CgSCR. Simulation results are consistent with the theoretical analysis in Section III.B.

*3) Verification of Critical Operating Conditions.* As discussed in Section III.A, when STATCOMs decrease CgSCR, under *critical* operating conditions that $gSCR=\lambda_{\min}\left\{\mathbf{S}_B^{-1}\mathbf{B}_{redn}\right\}$ in Eq. (8) and $I_{q\Sigma}$ in [-1,1], the *critical* subsystem tends to be most unstable. This is because when STATCOMs decrease CgSCR, the decrease of IBRs' active power output increases gSCR but decreases CgSCR, resulting in the increase of (gSCR-CgSCR). To verify *critical* operating conditions, we increase all IBRs' active power outputs in the modified 39-node system from 0.5 to 1 p.u. and set $I_{q\Sigma}$ of the corresponding *critical* subsystem as a constant value -0.5 p.u. The corresponding gSCR changes from 3.36 to 1.68. Fig. 7 shows the loci of dominant eigenvalues of the corresponding *critical* subsystem. We can see from scenario 1) in TABLE IV that when $I_{q\Sigma}$=-0.5 p.u., STATCOMs decrease CgSCR. Besides, we can see from Fig. 7 that with the increase of IBRs' active power outputs and $I_{q\Sigma}$=-0.5 p.u., the system becomes more unstable. Simulation results are consistent with theoretical analysis.

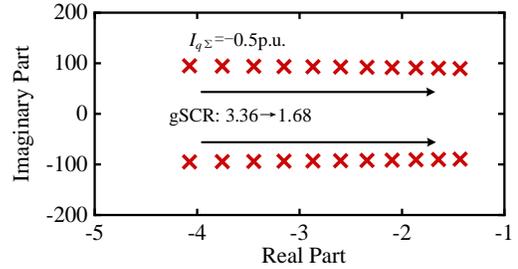

Fig. 7 Loci of dominant eigenvalues of the *critical* subsystem, when gSCR changes from 3.36 to 1.68 and $I_{q\Sigma}$ = -0.5 p.u..

*4) Verification that System's Stability is Bounded.* As discussed in Section IV.B, the small-signal stability of the multi-IBR system with STATCOMs under varying operating conditions is bounded by that of subsystem$_{1,1}$ and subsystem$_{1,n}$, which correspond to the most unstable and most stable subsystems. To verify this conclusion, we consider the scenarios in *1)* of this subsection. Fig. 5 (b) shows the loci of damping ratios of subsystem$_{1,1}$ and subsystem$_{1,9}$ for the modified 39-node system, which is represented by blue and grey dotted curves. We can see from Fig. 5 (b) that the red dotted curve is between blue and grey dotted curves. This indicates that the modified 39-node system is bounded.

*B. Verification of Adaptive Control Parameter Design Method*

The above subsection verifies that if STATCOMs decrease CgSCR, the system is most unstable under *critical* operating conditions; besides, the small-signal stability of the multi-IBR system with STATCOMs is bounded by subsystem$_{1,1}$~ subsystem$_{1,n}$. Therefore, the robust small-signal stability issues of the multi-IBR system with STATCOMs under varying operating conditions can be converted to that of subsystem$_{1,1}$~ subsystem$_{1,n}$ under *critical* operating conditions, with pre-condition that STATCOMs decrease CgSCR. Here, we furtherly verify the validity of the proposed control method for STATCOMs. Six cases are considered:

1) Divide *critical* operating conditions into 20 intervals;
2) Divide *critical* operating conditions into 10 intervals;
3) Divide *critical* operating conditions into 8 intervals;
4) Divide *critical* operating conditions into 4 intervals;
5) Divide *critical* operating conditions into 2 intervals;
6) Do not divide *critical* operating conditions.

Optimization problems Eq. (20) for cases 1)~6) are solved by Matlab solvers. The results for cases 1)~4) are given in TABLEs. V~VIII in Appendix. But optimization problems in Eq. (20) for cases 5)~6) have no solutions. This indicates that it may be hard to find fixed control parameters for STATCOMs to ensure robust small-signal stability of the multi-IBR system with STATCOMs under all operating conditions. Therefore, it is necessary to adaptively adjust STATCOMs' control parameters considering varying operating conditions.

Fig. 8 shows the CgSCR of the *critical* subsystem under *critical* operating conditions for cases 1)~4), and reference case that STATCOMs' control parameters refer to TABLE III, but PLL's PI parameters are set as 22, 7300. Under these five cases, the system's CgSCR is denoted as CgSCR$_{20}$, CgSCR$_{10}$, CgSCR$_{8}$, and CgSCR$_{0}$, respectively. Besides, gSCR and CgSCR of the modified 39-node system without STATCOMs under rated operating condition is 1.68 and 1.94.

As shown in Fig. 8, for the reference case, CgSCR may be larger (or smaller) than 1.94, i.e., STATCOMs may deteriorate (or improve) system stability. Moreover, increasing $I_{q\Sigma}$ may cause CgSCR>gSCR=1.68, i.e., the system becomes unstable. This demonstrates the necessity to dynamically adjust STATCOMs' control parameters to ensure the system's robust small-signal stability under varying operating conditions. In comparison, CgSCR of the *critical* subsystem under cases 1)~4) are always smaller than 1.5 (smaller than 1.68 and 1.94), when $I_{q\Sigma}$ is in the range of [-1,1]. That is, the obtained control parameters set for STATCOMs can ensure robust small-signal stability of the *critical* subsystem and that STATCOMs decrease the CgSCR of the *critical* subsystem under *critical* operating conditions. This illustrates that the proposed adaptive control parameter design method for STATCOMs can ensure robust small-signal stability of the multi-IBR system with STATCOMs under varying operating conditions. Besides, we can see that red and blue curves are commonly under brown and grey curves. It demonstrates that when the divided sub-conditions are more, the obtained STATCOMs' control-parameter set can improve the system's stability better.

Furthurly, the electromagnetic time-domain simulation of the modified 39-node system is provided to verify the proposed control method. Two operating conditions are considered: 1) all IBRs output rated active power and $I_{q\Sigma}$=-0.241 p.u.; 2) all IBRs output rated active power and $I_{q\Sigma}$=0.19 p.u. For each operating condition, we consider five scenarios:

I) All STATCOMs use control parameters in TABLE V;

II) All STATCOMs use control parameters in TABLE VI;
III) All STATCOMs use control parameters in TABLE VII;
IV) All STATCOMs use control parameters in TABLE VIII;
V) All STATCOMs use control parameters in TABLE III, but PLL's PI parameters are set as 22, 7300, as a reference.

For each scenario, a disturbance is applied to infinite bus at $t=1s$ to cause 0.05 p.u. voltage drops and lasts 0.05s. Time-domain responses of active power outputs of IBR1 under these scenarios are given in Fig. 9. In Fig. 9 (b), since control parameters of STATCOMs in TABLEs V~VIII in Appendix are the same when $I_{q\Sigma}$=0.19 p.u., we only provide responses of IBR1's active power output under scenarios IV) and V). We can see from Fig. 9 that the system is robustly stable under different operating conditions, when STATCOMs use proposed control method; but the system changes from stability to instability with fixed control parameters, when $I_{q\Sigma}$ changes from -0.241 to 0.19 p.u.. Simulation results are consistent with the analysis by Fig. 8 and thus verify the validity of the proposed control method for STATCOMs.

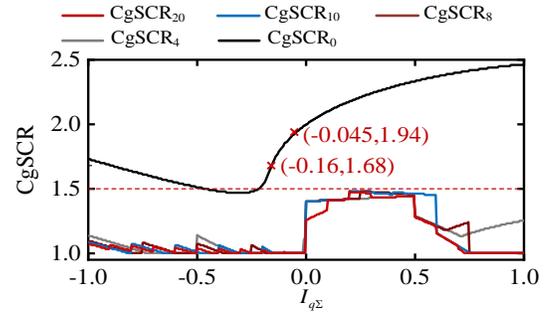

Fig. 8 CgSCR for the *critical* subsystem under *critical* operating conditions for cases 1)~4) and the reference case that control parameters of STATCOMs refer to TABLE III, but PLL's PI parameters are 22, 7300.

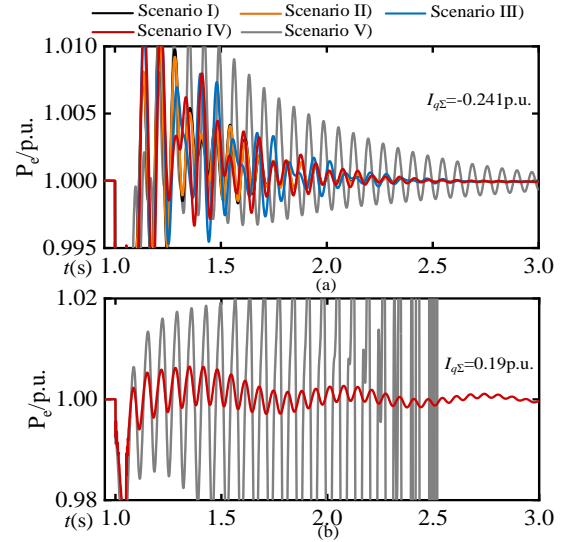

Fig. 9 Time-domain responses of IBR1's active power output under different scenarios for two operating conditions that all IBRs output rated active power: (a) $I_{q\Sigma}$=-0.241p.u.; (b) $I_{q\Sigma}$=0.19p.u..

## VII. CONCLUSIONS

This paper proposed an adaptive control-parameter design method for STATCOMs to ensure robust small-signal stability of a multi-IBR system with STATCOMs under varying operating conditions. It has been demonstrated that:

1) The proposed gSCR-based method can be used to evaluate the system's small-signal stability margin and to understand the impact mechanism of STATCOMs on IBR-induced oscillation issues under varying operating conditions from the viewpoint of grid strength.

2) The proposed control method for STATCOMs can ensure robust small-signal stability of the system under varying operating conditions, which avoids traversing all operating conditions and establishing the system's detailed models.

In future work, we will explore how to coordinate IBRs and STATCOMs with other devices (e.g., energy storage devices and static var capacitors, SVC) to ensure robust small-signal stability under varying operating conditions.

## APPENDIX

### TABLE V
#### OPTIMAL CONTROL-PARAMETER SET FOR STATCOMS UNDER CASE 1)

| $I_{gc}$ | [-1, -0.9) | [-0.9, -0.8) | [-0.8, -0.7) | [-0.7, -0.6) | [-0.6, -0.5) |
|---|---|---|---|---|---|
| PLL | 30.6+5796.4/s | 28.5+5857.9/s | 26.5+5915.0/s | 24.4+5966.5/s | 22.3+6009.7/s |
| AVC | 2.75+5/s | 2.77+5/s | 2.79+5/s | 2.81+5/s | 2.83+5/s |
| $I_{gc}$ | [-0.5, -0.4) | [-0.4, -0.3) | [-0.3, -0.2) | [-0.2, -0.1) | [-0.1, 0) |
| PLL | 20.2+6040.5/s | 18+6053.7/s | 15.5+6020/s | 12.8+6020/s | 9.6+5547.3/s |
| AVC | 2.85+5/s | 2.88+5/s | 2.91+5/s | 2.91+5/s | 2.98+5/s |
| $I_{gc}$ | [0, 0.1) | [0.1, 0.2) | [0.2, 0.3) | [0.3, 0.4) | [0.4, 0.5) |
| PLL | 5.1+3974.1/s | 1+100/s | 119.6+100/s | 1+100/s | 1+100/s |
| AVC | 3.01+5/s | 3.02+5/s | 3.02+5/s | 3.01+5/s | 3.01+5/s |
| $I_{gc}$ | [0.5, 0.6) | [0.6, 0.7) | [0.7, 0.8) | [0.8, 0.9) | [0.9, 1] |
| PLL | 10.3+20000/s | 9.2+14771.8/s | 10+13302.9/s | 11+12837.2/s | 12.1+12582.9/s |
| AVC | 2.93+5/s | 2.86+5/s | 2.79+5/s | 2.71+5/s | 2.64+5/s |

### TABLE VI
#### OPTIMAL CONTROL-PARAMETER SET FOR STATCOMS UNDER CASE 2)

| $I_{gc}$ | [-1, -0.8) | [-0.8, -0.6) | [-0.6, -0.4) | [-0.4, -0.2) | [-0.2,0) |
|---|---|---|---|---|---|
| PLL | 29.5+5791.6/ | 25.3+5902.2/ | 21+5974.7/ | 16.3+5942.2/s | 10.3+5381.8/s |
| AVC | 2.79+5/s | 2.82+5/s | 2.86+5/s | 2.92+5/s | 2.99+5/s |
| $I_{gc}$ | [0, 0.2) | [0.2, 0.4) | [0.4, 0.6) | [0.6, 0.8) | [0.8, 1] |
| PLL | 1+100/s | 119.6+100/s | 97.8+100/ | 9.2+14726.5/ | 11.1+12831.5/ |
| AVC | 3.02+5/s | 3+5/s | 2.97+5/s | 2.85+5/s | 2.7+5/s |

### TABLE VII
#### OPTIMAL CONTROL-PARAMETER SET FOR STATCOMS UNDER CASE 3)

| $I_{gc}$ | [-1, -0.75) | [-0.75, -0.5) | [-0.5, -0.25) | [-0.25, 0) | [0, 0.25) |
|---|---|---|---|---|---|
| PLL | 28.9+5785.7/s | 23.6+5909.5/s | 17.9+5927.3/s | 10.5+5316.3/s | 1+100/s |
| AVC | 2.8+5/s | 2.84+5/s | 2.91+5/s | 2.99+5/s | 3.02+5/s |
| $I_{gc}$ | [0.25, 0.5) | [0.5, 0.75) | [0.75, 1] | | |
| PLL | 105.2+100/s | 10.3+20000/s | 10.5+13036/s | | |
| AVC | 2.98+5/s | 2.92+5/s | 2.74+5/s | | |

### TABLE VIII
#### OPTIMAL CONTROL-PARAMETER SET FOR STATCOMS UNDER CASE 4)

| $I_{gc}$ | [-1, -0.5) | [-0.5, 0) | [0, 0.5) | [0.5, 1] |
|---|---|---|---|---|
| PLL | 25.8+5701.1/s | 12.2+4900.7/s | 1+100/s | 10.9+20000/s |
| AVC | 2.87+5/s | 3+5/s | 3.02+5/s | 2.9+5/s |